# Anyonic Braiding in a Chiral Mach-Zehnder Interferometer

Bikash Ghosh*,[1], Maria Labendik*,[1], Liliia Musina[1], Vladimir Umansky[1],

Moty Heiblum[1, #], and David F. Mross[2]

[1] Braun Center for Submicron Research & Department of Condensed Matter Physics

[2] Department of Condensed Matter Physics, Weizmann Institute of Science, Israel

* Equal contributions

# Corresponding author

#Corresponding author. Email: **moty.heiblum@weizmann.ac.il**

**Fractional quantum statistics are the defining characteristic of anyons. Measuring the phase generated by an exchange of anyons is challenging, as standard interferometry setups -- such as the Fabry-Pérot interferometer -- suffer from charging effects that obscure the interference signal. Here, we present the observation of anyonic interference and exchange phases in an optical-like Mach-Zehnder interferometer based on co-propagating interface modes. By avoiding backscattering and deleterious charging effects, this setup enables pristine and robust Aharonov-Bohm interference without any phase slips. At various fractional filling factors, the observed flux periodicities agree with the fundamental fractionally charged excitations that correspond to Jain states and depend only on the bulk topological order. To probe anyonic statistics, we use a small, charged top-gate in the interferometer bulk to induce localized quasiparticles without modifying the Aharonov-Bohm phase; however, with introducing periodic phase slips. The magnitude of the observed phase slips and their signs align with the expected value at filling 1/3, but their direction shows systematic deviations at fillings 2/5 and 3/7. Control over added individual quasiparticles in this design is essential for measuring the coveted non-Abelian statistics in the future.**

Quantum Hall states are quintessential examples of topological phases of matter ([1]). In particular, the fractional states host quasiparticle (QP) excitations that carry fractional charges ([2-4]) and exhibit anyonic statistics ([5-8]). Upon exchanging two identical QPs, the phase of their joint wavefunction changes by a fraction of π, unlike the behavior of fermions (bosons), where the exchange phase is π (2π) ([9]). Two key methods for probing these anyonic statistics involve interference in space ([10](#)) or in time ([11](#), [12](#)). In the spatial domain, the phase difference between two QP trajectories that enclose a magnetic flux is the Aharonov-Bohm (AB) phase. The number of the enclosed QPs determines an added statistical phase. For bosons or fermions, the statistical phase is an unobservable multiple of $2\pi$, but it becomes measurable in systems hosting anyons ([13](#), [14](#)). QPs can either be localized in the bulk (with a significant excitation energy) or delocalized in chiral edge modes at the periphery of the bulk (with a smaller excitation energy).

To probe these phases experimentally, interference studies of fractional quantum Hall (FQH) states utilize two primary architectures: the Fabry-Perot Interferometer (FPI) ([10](#), [15-19](#)) and the Mach-Zehnder Interferometer (MZI) ([20-22](#)). In the FPI, a quantum Hall bulk is bound on both sides by regions of a lower filling factor (Figure 1(c)). The interference loop consists of counter-propagating charge modes on the opposite sides of the bulk's central area. The loop is partially closed by two quantum point contacts (QPCs) that only allow tunneling of specific charges whose quantization is set by the bulk filling. Figure 1(c) illustrates the essential features of the FPI for the integer case using semiclassical electron trajectories. In particular, closed paths within the central region correspond to localized QPs within the interference loop. Promoting electrons to composite fermions (CFs) results in localized fractional QPs. Excitation of the latter changes the phase of the interfering QPs loops in discrete steps, also known as phase slips. In reality, the transmission through the FPI is additionally influenced by a `breathing' of the AB area with the magnetic field and, most importantly, by intrinsic charging energy for adding QPs to its central region ([23-27](#)). Still, sufficient screening of the Coulomb interaction allowed observation of phase slips, which approximately agree with the expected braiding phases ([26-29](#)).

In contrast, the conventional MZI contains an internal grounded drain, which renders it free of charging effects. However, this drain also makes it subject to the Byers-Yang argument ([30](#)), which predicts **$1\Phi_0$** periodicity ([31](#)). Physically, this periodicity arises because QPs returning to the internal loop continuously 'dress' the trivial AB interference of fractional charges by their braiding phase ([32-34](#)). Phase slips are neither observed ([34](#)) nor expected in the MZI ([32](#)); however, the visibility profile (as a function of the MZI transmission) drastically differs from that in the integer regime, consistent with its origin in anyonic interference ([34](#)).

An Optical-like Mach-Zehnder Interferometer (OMZI, read as OM-ZI) combines the MZI's immunity to charging effects with the FPI's ability to observe significant AB (super-) periodicities and anyonic braiding. An early version of the OMZI was first proposed by Giovannetti et al. in 2008 ([35](#)) and tested by Deviatov et al. in 2012 ([36](#)). A theoretical study by Batra et al. ([37](#)) presented an exact solution for specific edge modes interference (37). The OMZI structure is drawn schematically in Figure 1(a). It comprises three regions, each with a different filling factor. The central (bulk) region is tuned to filling $\nu_b$ while being bounded by two regions that are 'gate-tuned' to an upper filling $\nu_u>\nu_b$, and to a lower filling $\nu_l<\nu_b$. This configuration results in two co-propagating interface charged modes with corresponding 'fillings': $\nu_u-\nu_b$ and $\nu_b-\nu_l$ (38). Current emanating from source S1 (or source S2) is partitioned by the input QPC1 to two forward-propagating chiral interface modes surrounding the bulk. The two interface modes are re-partitioned by the output QPC2, and the current is eventually collected by drain D1 (or drain D2). The chirality of the charged modes ensures that all the sourced current passes through the OMZI and is not subject to charging effects. Additionally, the potential landscape, which exhibits a saddle instead of a valley in the central region, implies the absence of localized low-energy QPs inside the interference loop (Figure 1(c)). These two features distinguish the OMZI from the FPI and allowed it to exhibit pristine AB interference without phase slips.

To observe anyonic braiding, the QPs must be introduced into the OMZI by a separate mechanism. For this purpose, we employ a top gate (TG) that locally partially depletes (accumulates) charge upon application of a negative (positive) gate voltage. A reduced filling factor under the TG leads to an interface mode in its periphery, which supports fractionally quantized charged excitations (encoded as the winding number of the interface mode). These QPs have lower energies than bulk QPs and thus can be pushed across the Fermi level by varying the TG voltage or the magnetic field. Thereupon, fractional charge transfers from the gapless outer boundary of the OMZI to the inner edge mode encircling the small, gated region. The resulting QPs in the center of the OMZI are spatially isolated from the interfering QPs (at the far edge) and affect them solely via the quantum statistical phases.

Our OMZI (Figures 1(a) & 1(b)) is fabricated in a 2DEG, induced in GaAs-AlGaAs heterostructure, with a density of $1.07 \times 10^{11} cm^{-2}$ and 'dark mobility' $4.5 \times 10^6 cm^2/V\text{-}S$ at 4.2K. Measurements are performed at the electrons' temperature of ~15mK. The fabricated internal area of the OMZI is $A=3\mu m^2$, with a single path length of 3μm. Due to the gates' depletion, the actual OMZI area is smaller and affected by the gates' voltage and that of the two QPCs. The isolated TG, with a fabricated area $A_{TG}=0.785\mu m^2$, is charged via an air bridge (Figure 1(b)). A small AC signal (~$2\mu V_{RMS}$), at a frequency of 776kHz, is applied to source S1. The interfering signal reaches the two drains, with the signal monitored in D1. The signal is filtered by an LC circuit (with a BW=30kHz) and subsequently amplified by an amplifier cooled to 4.2K (Gain=10), followed by a room-temperature amplifier (Gain=400). A spectrum analyzer, tuned to the input carrier frequency, averages the amplified output signal with a narrow bandwidth, BW=5Hz.

We tested three configurations with distinct fractional bulk fillings: $\nu_b$=1/3, $\nu_b$=2/5, and $\nu_b$=3/7. The upper and lower gated regions were selected to realize two co-propagating interface modes formed in the following configurations: $\nu_u$-$\nu_b$-$\nu_l$ = 2/3-1/3-0, and 1-2/5-0, and 1-3/7-0. Upstream neutral modes at the upper positively gated interface accompany the co-propagating interface charge modes (see Figure 1 and schematic in S11), expected to suppress the visibility (indeed found to be 2-10%). The two OMZI's QPCs were tuned to partition the innermost interface mode in all three configurations, i.e., the one carrying 1/3, 1/15, and 1/35 (in units of the quantum conductance), with an average transmission kept at $t_{S1-D1}$=0.8 at $\nu_b$=1/3 and $t_{S1-D1}$=0.5 at $\nu_b$=2/5 and 3/7; all with $V_{TG}$=0. The flux periodicities in the three configurations matched the expectations based on the partitioned fractional charges at the two QPCs (Figure 2). The actual inner area of the OMZI is not known accurately (being depleted near the $\nu$=0 side and expanded near the accumulated side), yet it is also dependent on $\nu_b$; likely smaller than the fabricated area (by some ~5-10%). For $\nu_b$=1/3 the observed periodicity matches the expected value ~3$\Phi_0$, with the actual OMZI area taken to be 2.6 µm². Similarly, at $\nu_b$=2/5, $\nu_b$=3/7, we find ~5$\Phi_0$ (actual area 2.76 µm²), and ~7$\Phi_0$ (actual area 3.2 µm²). The periodicities in $V_{MG}$ obey B×$\Delta V_{MG}^i$=85±3 Tesla-mV. As anticipated, for $V_{TG}$=0, the B-$V_{MG}$ pajamas are free of phase flips. It is worth noting that interference of integer modes, tested in configurations, 2-1-0, 3-2-0, and 4-2-0 (see Supplementary S5), was unsuccessful due to a lack of partitioning in the QPCs, likely because of opposite spins of the co-propagating integer edge modes.

The visibility of the interference signal is defined as $\upsilon_e = \frac{t_{max}-t_{min}}{t_{max}+t_{min}}$, where $t_{max}$ and $t_{min}$ are the maximum and minimum transmissions at each drain (peak and valley of the AB oscillations), respectively. The observed visibilities varied around 2-5% at an average transmission of $t_{S1-D1}$=0.8 for $\nu_b$=1/3 and $t_{S1-D1}$=0.5 for $\nu_b$=2/5 and 3/7 (inner modes at higher fillings). The visibilities of the outer modes of $\nu_b$=2/5 and 3/7 are comparable to those of the inner modes. We attribute the significantly smaller visibilities in the OMZI compared to a conventional MZI ([20]) to the topological neutral modes at the interface (see Supplementary S10).

The theoretically expected braiding phases for Jain states have been obtained in several studies ([7, 9, 13, 27, 39, 40]). These incompressible FQH states arise at fractional fillings, $\nu = \frac{n}{2pn+1}$, with the integer n denoting the number of filled composite fermion (CF) levels and p the CF generation. Here, p=1 corresponds to two-flux quanta attached to the electrons. Elementary QPs of the Jain states carry a fractional charge, $e^* = \frac{e}{2pn+1}$ and obey fractional statistics. Exchanging two QPs yields a phase factor $\pm \frac{2\pi p}{2np+1}$, depending on the orientation of the exchange path. The two interfering trajectories in the OMZI differ by a full loop around the interferometer center and are expected to acquire a quantized phase difference of $\frac{\theta^*}{2\pi} = \frac{2p}{2np+1}$ for each localized QP(39). Here, we consider the fillings, $\frac{1}{3}, \frac{2}{5}, \frac{3}{7}$, with the expected braiding phases $\frac{\theta^*}{2\pi} = \frac{2}{3}, \frac{2}{5}, \frac{2}{7}$ (see Supplementary S2).

To observe these braiding phases, we studied the AB pajamas at a range of TG voltage, $V_{TG}$=-(50-100)mV for the three bulk filling configurations, with the two QPCs tuned to partition the **innermost** edge mode in each of the three configurations. The observed quantized phase slips (Figure 3) occur approximately when the magnetic flux piercing the partly depleted area under the TG changes by $\mathbf{1\Phi_0}$ (independent of the modulating gate voltage, $V_{MG}$). The phase slip magnitudes determined either directly from the 2D pajama plots or via the 'lock in' procedure (see Methods Section and Supplementary S3) generally agree with the (absolute) expected phase slips (Table 1 in S2). Studies of the interfering partitioned outer modes in the QPCs' and the corresponding phase slips are presented in Supplementary S11. We note that the signs of some phase slip at filling $\nu = \frac{2}{5}$ and $\nu = \frac{3}{7}$ are opposite to the theoretical expectations. An increasing magnetic field reduces the filling factor, which is expected to create quasi-holes inside the interferometer. Instead, the observed phase jumps at $\nu = \frac{2}{5}$ correspond to added QPs. At the filling factor $\nu = \frac{3}{7}$, both directions of jumps are observed. A possible explanation of the unexpected phase-slip signs could arise from the energetics under the top gate, which may favor a simultaneous introduction of more than one quasi-hole (or quasi-particle). For example, four e/5 quasi-holes at $\nu$=2/5 are equivalent to one integer hole and one e/5 quasi-particle. The interference phase is only sensitive to the fractional charge and would thus exhibit phase slip associated with one added quasi-particle under the top gate.

The most striking observation of anyonic braiding in the OMZI occurs when B and $V_{MG}$ are held constant, allowing the AB phase to remain unchanged while varying $V_{TG}$. This process introduces QPs into the bulk's center without disturbing the interference pattern, enabling precise detection of phase slips. This test was performed in the configurations $\nu_b$=1/3 and $\nu_b$=2/5. A 2D plot of the OMZI conductance as a function of $V_{TG}$ and B is shown in Figure 4 (being a different pajama). At a constant magnetic field near a conductance step, a minute voltage change applied to the TG ($\Delta V_{TG}$~1mV) leads to a sharp jump in the conductance of the OMZI (Figure 4). To translate the conductance steps to a phase slip in the AB interference, Figures 4(b) & 4(d) show the AB oscillations (as a function of B, on both sides of vertical dashed lines that mark the conductance steps). The abrupt shift in the AB oscillation (across $\Delta V_{TG}$=1mV) directly reveals the quantized phase slip due to added QPs at the TG. The detailed size of the phase slips is provided in the captions of Figure 4.

Our experiments utilized three different modes of the newly developed OMZI: **(i)** Without charging the Top Gate (TG), we find pristine AB oscillations with flux periodicities tied to the QP's fundamental charge. **(ii)** With a charged TG ($V_{TG} \neq 0$), localized QPs are introduced periodically with a changing magnetic field. Consequently, the OMZI shows phase slips corresponding to the QP's braiding phase in the Aharonov Bohm (AB) interference pattern. **(iii)** At a constant magnetic field and a constant modulation gate voltage ($V_{MG}$), localized bulk QPs are introduced one-by-one with $V_{TG}$, leading to abrupt slips in the AB phase - as is expected due to QP braiding.

Similar tests performed with different co-propagating 'interface modes' proved that bulk filling - and not the interface modes - determines the interfering charges and their exchange phase. The versatility of the OMZI geometry requires minor modifications to realize different probes of QP's statistics. For example, time braiding ([11], [12]) can be performed using the first QPC to dilute the QP current, which will braid with thermally excited QPs in the second QPC. Attaching two OMZIs will form a Hanbury Brown and Twiss (HBT) configuration ([22])([35]), allowing intensity interference of QPs arriving from two different sources. In addition, it will be easy to perform Hong, Ou, Mandel (HOM) experiments on QPs ([41]). Moreover, the OMZI is an ideal interferometer for obtaining interference and determining the topological order of non-Abelian FQH states.

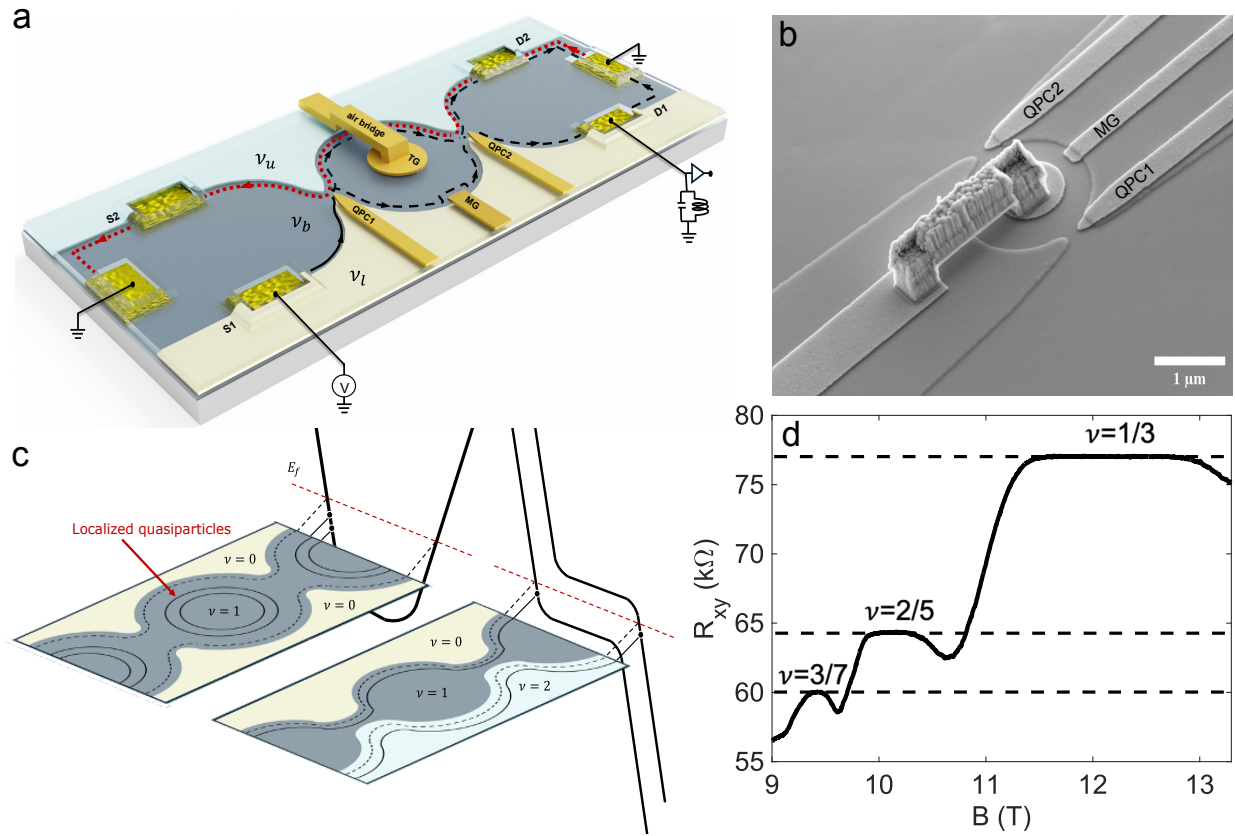

**Figure 1. Optical-like Mach Zehnder interferometer (OMZI) and Fabry Perot interferometer (FPI).**
**(a)** A schematic of the OMZI. The lower gate (light yellow with filling factor $\nu_l$) is biased negatively, depleting the density. The upper gate (light blue with filling factor $\nu_u$) is biased positively, increasing the density. Co-propagating modes (downstream in black) are partitioned at QPC1 and QPC2. The source S1 is charged and the interfering signal is measured in D1. Upstream neutral modes are drawn in dotted red lines. An anti-dot in the middle of the OMZI bulk is formed by a negatively charged top gate (TG) connected via an air bridge to a power supply. **(b)** An SEM micrograph of the OMZI. **(c)** A comparison between the FPI and the OMZI for integer quantum Hall states based on semiclassical electron trajectories. In the FPI (left) states below the Fermi level $E_f$ form closed orbits corresponding to localized quasiparticles (QPs). Fully or partially closed trajectories suffer from Coulomb interactions. In the OMZI (right), all trajectories extend beyond the central region and are insensitive to charging effects. **(d)** The Hall resistance is in the range of the fractional states under study.

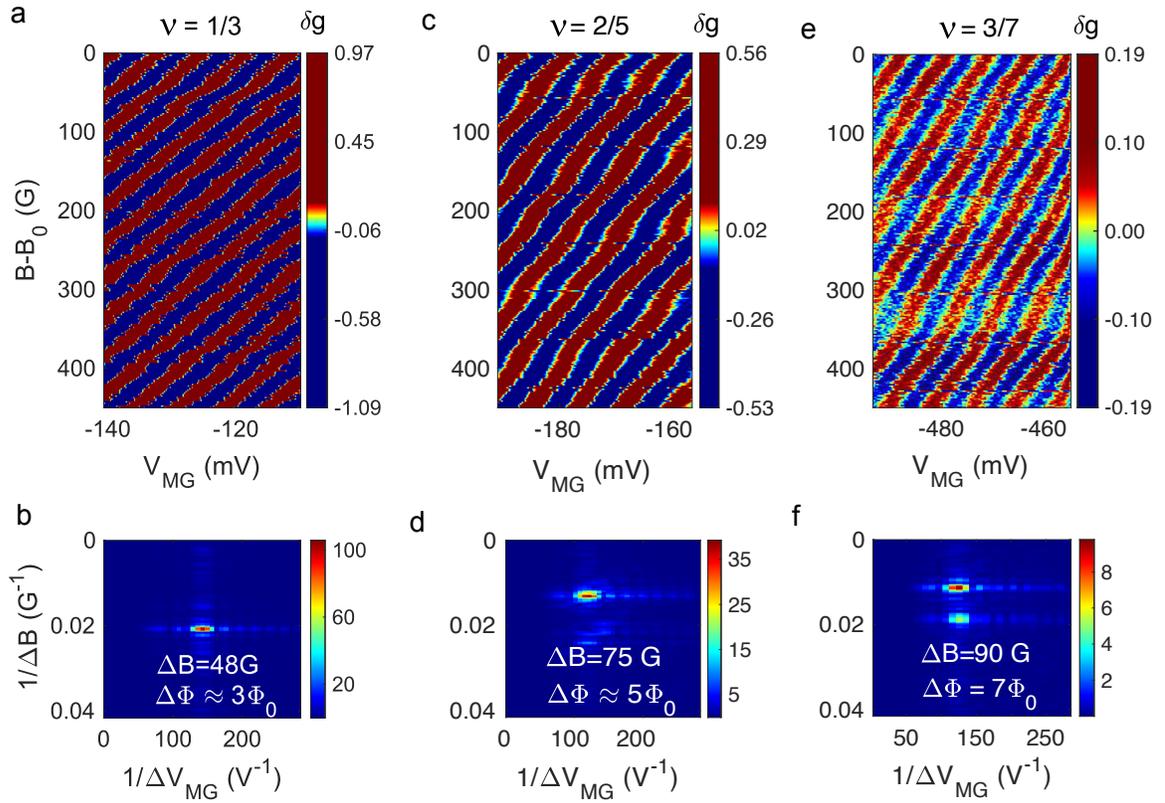

**Figure 2. Aharonov Bohm (AB) pajamas plot of co-propagating edge modes in an Optical-like Mach-Zehnder Interferometer (OMZI).** The 2D AB interference as a function of magnetic field B and the modulation-gate voltage $V_{MG}$. The pajamas in **a**, **c**, and **e** correspond to partitioning of the innermost modes such that the interfering quasiparticles (QPs) carry (w/charges: e/3, e/5, and e/7). The FFTs in **b**, **d**, and **f** reflect the expected periodicities. $\Delta\Phi = (e/e^*)\Phi_0$. Phase slips related to QP excitations are absent, as expected.

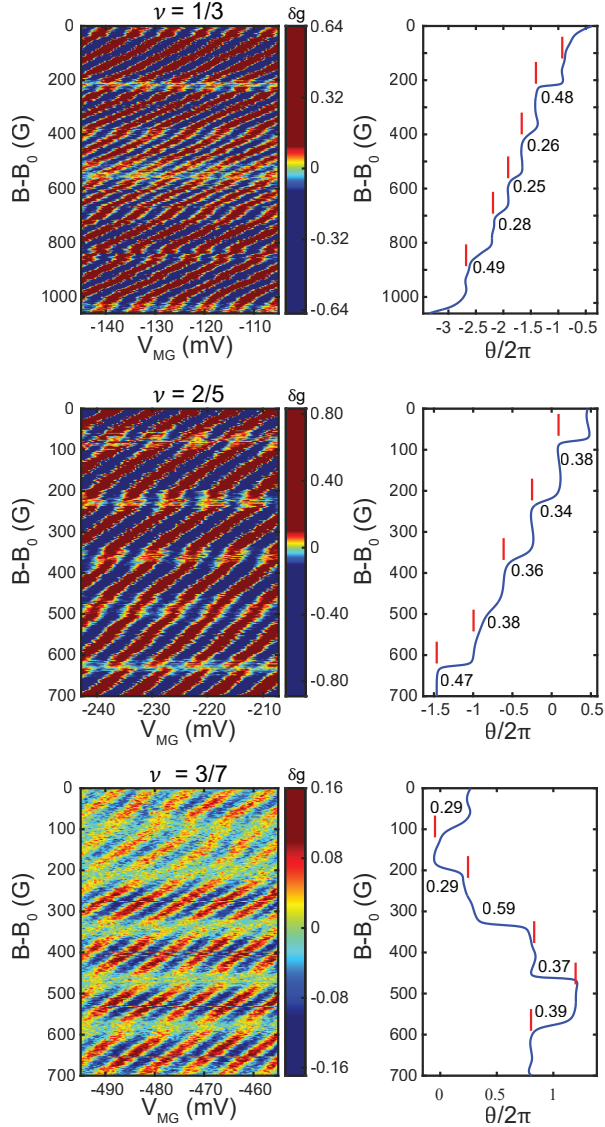

**Figure 3. Aharonov Bohm (AB) pajamas with magnetic field excitations of local quasiparticles**.

Panels **a, c,** and **e** show the 2D AB interference as a function of magnetic field B and the modulation-gate voltage $V_{MG}$. Clear phase slips appear for approximately every flux quantum piercing the 'effective area' under the TG. The top gate (TG) of the anti-dot is charged with a voltage in the range of -50mV to -100mV (at $\nu_b$=1/3, $V_{TG}$=-90mV; at $\nu_b$=2/5, $V_{TG}$=-50mV; and at $\nu_b$=3/7, $V_{TG}$=-70mV), leading to a weak depletion under the gate. he fabricated area of the TG is 0.78µm$^2$; however, the effective area extracted from the distance between phase jumps is 0.27μm²@$\nu_b$=1/3; 0.3μm²@$\nu_b$=2/5, and 0.34μm²@$\nu_b$=3/7, suggesting a smaller depleted island under the TG. Panels **b, d,** and **f** show the phase slips calculated with the 'Lock-In technique' (see **Methods** and Supplementary S3). The vertical red lines indicate the value of the phase in regions where it is constant, and the values of phase shifts between two consecutive regions are indicated. The magnitudes of the phase slips (modulo 2π) generally agree with the theoretically expected value $\frac{\theta^*}{2\pi} = \frac{2p}{1+2np}$ (see **Table 1 in S2)**. At $\nu_b$=1/3, all phase jumps are consistent with the expected creation of quasi-holes as the magnetic field increases. Surprisingly, the jumps at $\nu_b$=2/5 instead correspond to inserted QPs. At $\nu_b = 3/7$, both types of phase jumps are present.

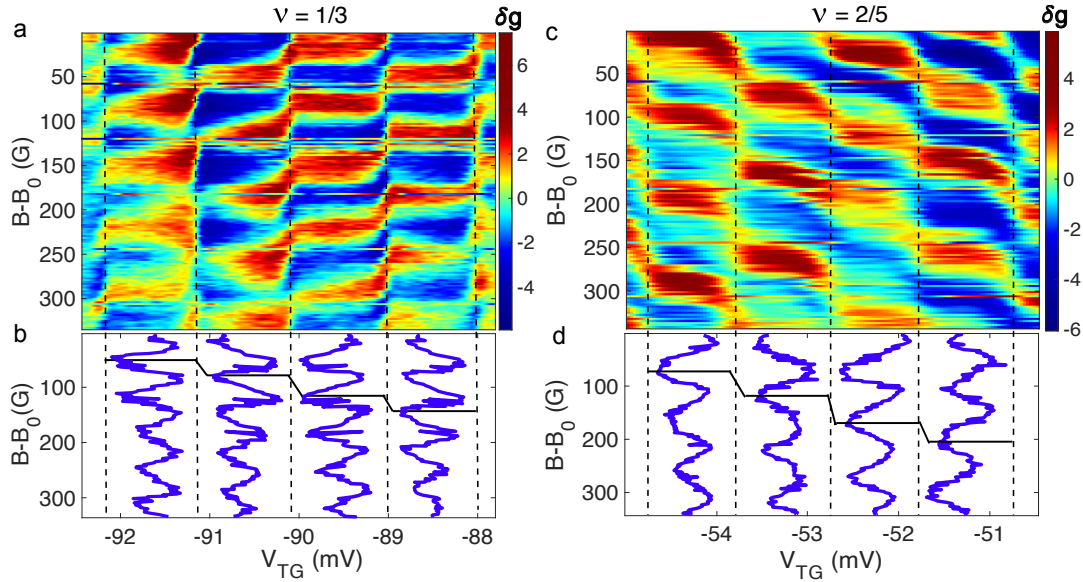

**Figure 4. Local excitation of quasiparticles by activating $V_{TG}$ the local anti-dot top gate (TG).**

2D plots of the AB interference for $\nu_b$=1/3 and $\nu_b$=2/5 of partitioned innermost edge modes as a function of the magnetic field, B, and the top gate voltage $V_{TG}$ (**a & c**). At any constant B, changing the applied voltage $V_{TG}$ by $\Delta V_{TG}$=1mV leads to sharp phase jumps indicated by the vertical dotted lines. Panels (**b & d**) show the conductance oscillations with the B for each region separated by phase jumps. Comparing two neighboring regions yields the actual phase difference. For $\nu_b$=1/3, the average magnitude of the jumps is $\frac{\theta^*}{2\pi} = 0.38$, while for $\nu_b$=2/5 it is $\frac{\theta^*}{2\pi} = 0.42$. At $\nu_b$=3/7, the visibility was too low to extract reliable values (see expected for $\theta^*$ @ Supplementary S2, **Table 1)**.

## Methods

**Sample fabrication process**

The MESA of the device, with dimensions 250×650 μm$^2$ was prepared in GaAs/AlGaAs heterostructure by wet etching in $H_2O_2$: $H_3PO_4$: $H_2O$ =1:1:50 for 100 S. The 2DEG depth was 170 nm from the surface. The ohmic contacts were deposited at the edge of the MESA as well as within the MESA (for the interface modes) in a sequence, Ni (15 nm), Au (260 nm), Ge (130 nm), Ni (87.5 nm), Au (15 nm) from the top GaAs surface. This process was followed by annealing at 440°C for 80 seconds. The sample was covered by 30nm of HfO$_2$ layer to electrically separate the ohmic contacts from the large metal gates (Ti (5 nm)/Au (15 nm)). Two large metal gates separated the MESA into three regions with filling factors $\nu_u$, $\nu_b$, $\nu_l$. In the following step, the MESA was covered again by 25nm HfO2 layer, followed by the deposition of QPCs, the modulation gate (MG), and the top gate (TG). In the final step, ohmic contacts, large metallic gates, QPCs, and MG were connected to pads via thick gold lines. The TG was connected by an air bridge in the interferometer region, followed by gold lines that passed over the HfO$_2$ that coated the metallic gates.

**Geometric phase analysis**

The phase slips are analyzed using the spatial lock-in method. Namely, the Aharonov-Bohm oscillations described by $T(\mathbf{r}) = A(\mathbf{r})\sin(\mathbf{qr} + \theta(\mathbf{r}))$, where $\mathbf{r} = (B, V)$, $\mathbf{q} = (q_B, q_V)$ is the frequency of the AB oscillations and the $\theta(\mathbf{r})$ is the phase shift due to the phase slips. To extract $\theta(\mathbf{r})$, two lock-in signals are defined $s_1(\mathbf{r}) = \cos(\mathbf{qr})$ and $s_2(\mathbf{r}) = \sin(\mathbf{qr})$. The original signal $T(\mathbf{r})$ is multiplied by these signals, and low-pass filtering is applied to obtain outputs proportional to $\sin\theta(\mathbf{r})$ and $\cos\theta(\mathbf{r})$. The ratio of these allows us to calculate the phase slips as $\theta(\mathbf{r}) = \arctan\frac{\text{LPF}(s_1(\mathbf{r})\times T(\mathbf{r}))}{\text{LPF}(s_2(\mathbf{r})\times T(\mathbf{r}))}$ (see Supplementary S3).

**Data availability**

The data that support the plots within this paper and other findings of this study are publicly available at https://doi.org/10.5281/zenodo.15394075

**Acknowledgments**
M.H. thanks Dima E. Feldman for the fruitful discussions and Mitali Banerjee for her suggestions. B.G and M.L thanks Arup Kumar Paul, H.K. Kundu and S. Biswas for the helpful comments that improved our device. B.G and M.L thank Ambikesh Gupta for the help with statistical phase analysis. D.F.M. acknowledges many illuminating conversations on quantum Hall interferometry with Yuval Ronen. D.F.M. was supported by the Israel Science Foundation (ISF) under Grant No. 2572/21 and by the Deutsche Forschungsgemeinschaft (DFG) within the CRC network TR 183 (project Grant No. 277101999). M.H. acknowledges the support of the European Research Council under the European Union's Horizon 2020 research and innovation program (Grant Agreement No. 833078). M.L. thanks the Ariane de Rothschild Women Doctoral Program for their support.

**Author contributions**
B.G. fabricated the devices. B.G. and M.L. performed the measurements and analyzed the data with the input from M.H. L.M. characterized the devices at the initial stage of the experiment. M.H. supervised the experiment's design, execution and data analysis. D.F.M. worked on the theoretical aspect and data analysis. V.U. grew the GaAs heterostructures. All authors contributed to the writing of the manuscript.

**Competing interests** The authors declare no competing interests.

# Supplementary Material

## S1. Quasiparticles in FPI and OMZI interferometers

Fractional Quantum Hall (FQH) interferometers probe bulk properties such as quasiparticle charges and their statistics by employing edge states. Quasiparticles in the bulk are well-defined, manifesting as local finite-energy excitations with a quantized charge. In contrast, the gapless edge modes permit excitations of arbitrary charge. To bridge the conceptual difference, adopting a shared language to describe both bulk and edge properties is helpful.

The boundary of any fractional quantum Hall state at filling factor $\nu$ supports a charge mode, i.e., a phase field $\phi$ governed by the Hamiltonian, $H_{\text{edge}} = u \int dx \, (\partial_x \phi)^2$. This mode encodes the charge density on the edge via, $\rho = \frac{1}{2\pi} \partial_x \phi$, in units of the electron charge. A *finite* quantum Hall droplet of electrons can only carry an integer charge. Consequently, the total charge on the edge, which is given by the winding of $\phi$ in units of $2\pi$, i.e., $Q = \frac{1}{2\pi} \int dx \, \partial_x \phi$ is an integer for any $\nu$, unless the droplet's bulk contains quasiparticles. In systems with multiple boundaries (e.g., inner and outer edges in the Corbino geometry), the total charge on all edges, $Q \equiv \sum_i Q_i$ is also quantized as an integer. By contrast, the charges $Q_i$ on individual edges can change in steps of the fundamental quasiparticle charge $e^*$.

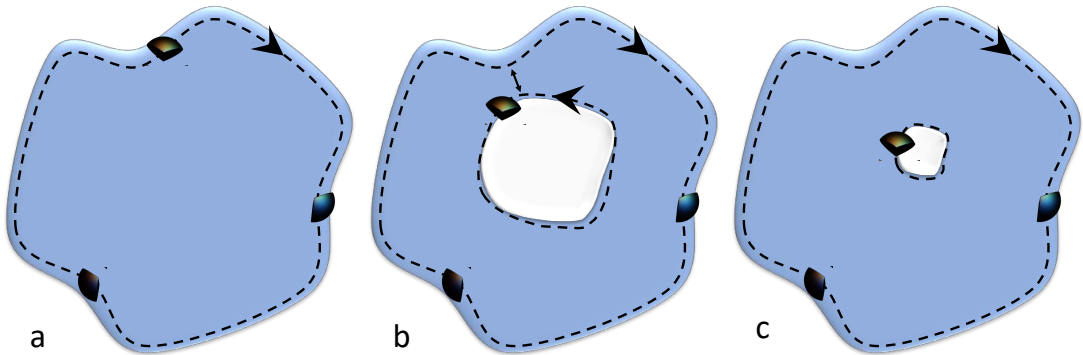

**Figure S1**: **Quasiparticles in a fractional quantum Hall droplet. (a)** The gapless edge of an FQH droplet supports arbitrarily charged excitations, provided the total charge is an integer. **(b)** Quasiparticles with fractional charge $e^*$ can transfer between the inner and the outer edge, similar to a quantum point contact (QPC). **(c)** Shrinking the inner region to a point when the inner edge carries fractional charge results in a bulk quasiparticle.

Bulk quasiparticles can be described as edge modes encircling a small region of altered filling factor. Consider a small 'anti-dot' region with zero filling, much smaller than the quantum Hall droplet. This system realizes the Corbino geometry, with an outer edge mode encircling the droplet and an inner edge around the anti-dot. The states on the inner edge are subject to finite size quantization, meaning their energy levels are discrete. Varying the magnetic field, electron density, or the anti-dot area continuously changes these energies. When it becomes favorable, a fractional charge $e^*$ transfers from the droplet's outer (longer) edge to the (shorter) periphery of the anti-dot. The fractional charge on the inner edge is equivalent to a bulk quasiparticle. Specifically, slowly shrinking a well-isolated anti-dot to a point does not change charge or statistics, which are quantized. Once the anti-dot has been fully removed, a bulk quasiparticle remains.

This analysis holds even when the anti-dot's filling factor is non-zero. If the anti-dot realizes a fractional quantum Hall state at a filling factor $v'$, the total charge in its vicinity is still quantized according to the surrounding bulk. If that charge had a different fractional value, shrinking the dot to zero would leave behind a finite energy excitation with a charge not supported by the quantum Hall bulk, creating a contradiction. In terms of the edge modes, the anti-dot at filling $v'$ creates an additional charge mode at its perimeter. Since the anti-dot only has a single (outer) boundary, this additional mode can only carry an integer charge (i.e., $2\pi$ windings), while the bulk's inner mode permits $2\pi e^*$ windings. The two combine into a total charge mode with $2\pi e^*$ windings at the perimeter of the anti-dot. **Therefore, the total charge localized near the anti-dot remains quantized at $e^*$, dictated by the surrounding bulk at filling $v$**. If the anti-dot supports fractional charges $e'^* \neq e^*$, the charges only exist as neutral 'quasiparticle—quasihole' pairs, which do not contribute to the braiding phase.

The central region of a Fabry Perot Interferometer (FPI) behaves similarly to an anti-dot, with the outside regions of the FPI acting as the bulk. Changes in the magnetic field or area affect the number of quasiparticles (edge-mode excitations) localized along the FPI's inner periphery, as shown by the closed loops in Fig. 1(c). However, unlike in an isolated anti-dot, the edge states couple to the states extending beyond the central part of the FPI, reducing their lifetime. The introduction of localized quasiparticles with a magnetic field and their coupling to the outside modes are intrinsic properties of FPIs and require careful treatment. By contrast, these effects are absent in both the standard MZI and the OMZI Interferometers.

In the OMZI, all charge modes co-propagate and do not support localized states in the central region [Fig. 1(c)]. This desirable feature comes at the cost of only displaying AB phases without anyonic braiding contributions (Fig. 2). To observe the latter; quasiparticles must be introduced through a different mechanism, such as a top-gate (TG) defining an anti-dot. Varying the voltage of the TG introduces quasiparticles, which transfer from the outer edge. Alternatively, changing the magnetic field alters the bulk filling factor and introduces quasiparticles that can either be delocalized on the outer edge or localized at the anti-dot. As a rule of thumb, localized particles are expected to appear whenever the magnetic flux through the anti-dot area changes by one flux quantum. Since this area is much smaller than the center of the interferometer, quasiparticles are introduced at a reduced rate compared to that in the FPI. Moreover, these quasiparticles are well isolated from the interfering modes.

## S2. Braiding phases of Abelian quasiparticles

The statistical properties of quasiparticles in Abelian quantum Hall states orders are efficiently captured using the K-matrix formalism **[1]**. An integer-valued matrix $K$ and a charge vector $\boldsymbol{t}$ encode a state at bulk filling factor $v = t^T K^{-1} t$, whose vacuum interface is described by the Lagrangian,

$$L = \frac{1}{4\pi} \sum_{i,j} \partial_x \phi_i (K_{ij} \partial_t \phi_j - V_{ij} \partial_x \phi_j),$$

with a non-universal velocity matrix $V$. All bulk quasiparticles of the topological order $(K, t)$ are specified by integer vectors $l$ with the electric charge given by

$$Q(l) = t^T K^{-1} l. \tag{1}$$

A quasiparticle defined by $l_1$, being fully encircled by another defined by $l_2$, yields the braiding phase,

$$\theta(l_1, l_2) = 2\pi \, l_1^T K^{-1} l_2. \tag{2}$$

For Jain states at of filling $v = \frac{n}{2pn+1}$, $K$ is an $n \times n$ matrix, which can be expressed using the identity matrix and matrix of ones as,

$$K = \begin{pmatrix} 1 & \cdots & 0 \\ \vdots & \ddots & \vdots \\ 0 & \cdots & 1 \end{pmatrix} + 2p \begin{pmatrix} 1 & \cdots & 1 \\ \vdots & \ddots & \vdots \\ 1 & \cdots & 1 \end{pmatrix}, \tag{3}$$

with $n$-component charge vector $t = (1, \ldots, 1)$. Inverting Eq. (3) yields,

$$K^{-1} = \begin{pmatrix} 1 & \cdots & 0 \\ \vdots & \ddots & \vdots \\ 0 & \cdots & 1 \end{pmatrix} - \frac{2p}{1+2np} \begin{pmatrix} 1 & \cdots & 1 \\ \vdots & \ddots & \vdots \\ 1 & \cdots & 1 \end{pmatrix}. \tag{4}$$

For a fundamental quasiparticle $l_0 = (1, 0, 0, \ldots)$, Eqs. (1) and (2) yield the known results for *Jain states* [1],

$$Q(l_0) = \frac{1}{1+2np},$$

$$\theta(l_0, l_0) = 2\pi - 2\pi \frac{2p}{1+2np}.$$

For the first Jain's states, $p = 1$. Ignoring the first $2\pi$ term, the reduced phase is: $\frac{\theta^*}{2\pi} = \frac{2}{1+2n}$ (see above).

[1] *Quantum Field Theory of Many-body Systems*, Xiao-Gang Wen (Oxford Graduate Text, 2004).

| ν | n | theory: $\frac{\theta^*}{2\pi}$ mod 1 | $\frac{\theta^*}{2\pi}$, from $B - V_{MG}$ | $\frac{\theta^*}{2\pi}$, from $B - V_{TG}$ |
|---|---|---|---|---|
| 1/3 | 1 | $-\frac{1}{3} = -0.33$ | $-0.35 \pm 0.11$ | $-0.38$ |
| 2/5 | 2 | $\frac{2}{5} = 0.40$ | $-0.39 \pm 0.05$ | $-0.42$ |
| 3/7 | 3 | $\frac{2}{7} = 0.29$ | $\pm 0.35 \pm 0.05$ | $-$ |

**Table 1**. Comparison of the theoretically expected value $\frac{\theta^*}{2\pi} = \frac{2}{1+2n}$ for the first three Jain fillings. Spatial lock-in method calculation of the 'pajamas' in Fig. 3 ($B - V_{MG}$), the average of the phase jumps is taken in the convention that all the phase jumps are in the range between $-\pi$ and $\pi$ ; direct phase jumps from the B dependence across the top gate abrupt $\Delta V_{TG}$ change in Fig. 4 ($B - V_{TG}$).

## S3. Phase slips analysis by spatial *lock-in method*

The main interference signal in the space of $B$ and $V_{MG}$ (and $V_{TG}$) is periodic and thus can be described by:

$$T(\mathbf{r}) = A(\mathbf{r}) \sin(\mathbf{qr} + \theta(\mathbf{r})), \qquad (1)$$

where $\mathbf{q} = (q_B, q_V)$ is the frequency of the Aharonov-Bohm oscillations, $\mathbf{r} = (B, V)$, $\theta(\mathbf{r})$ is the additional phase that remains constant in the absence of the phase-slips and is increasing stepwise after each phase slip by the value of the statistical phase.

To extract the statistical phase $\theta(r)$, we define the *lock-in signal*,

$$\begin{aligned} s_1(\mathbf{r}) &= \cos(\mathbf{qr}) \\ s_2(\mathbf{r}) &= \sin(\mathbf{qr}) \end{aligned} \qquad (2)$$

Multiplying the original signal $T(r)$ by $s_1$ and $s_2$ separately,

$$\begin{aligned} s_1(\mathbf{r}) \times T(\mathbf{r}) &= A(\mathbf{r})\cos(\mathbf{qr}) \sin(\mathbf{qr} + \theta(\mathbf{r})) \\ &= \frac{A(\mathbf{r})}{2}(\sin(\mathbf{qr} + \mathbf{qr} + \theta) + \sin(\mathbf{qr} + \theta - \mathbf{qr})) \\ &= \frac{A(\mathbf{r})}{2}(\sin(2\mathbf{qr} + \theta(\mathbf{r})) + \sin\theta(\mathbf{r})) \\ s_2(\mathbf{r}) \times T(\mathbf{r}) &= A(\mathbf{r})\sin(\mathbf{qr}) \sin(\mathbf{qr} + \theta(\mathbf{r})) \\ &= \frac{A(\mathbf{r})}{2}(\cos(\mathbf{qr} + \theta(\mathbf{r}) - \mathbf{qr}) - \cos(\mathbf{qr} + \theta(\mathbf{r}) + \mathbf{qr})) \\ &= \frac{A(\mathbf{r})}{2}(\cos\theta(\mathbf{r}) - \cos(2\mathbf{qr} + \theta(\mathbf{r}))) \end{aligned} \qquad (3)$$

Finally, adding low-pass filtering (LPF) to the expressions (3)

$$LPF(s_1(\mathbf{r}) \times T(\mathbf{r})) = \frac{A(\mathbf{r})}{2} \sin \theta(\mathbf{r})$$
$$LPF(s_2(\mathbf{r}) \times T(\mathbf{r})) = \frac{A(\mathbf{r})}{2} \cos \theta(\mathbf{r})$$
(4)

$$\frac{LPF(s_1(\mathbf{r}) \times T(\mathbf{r}))}{LPF(s_2(\mathbf{r}) \times T(\mathbf{r}))} = \tan \theta(\mathbf{r})$$
$$\theta(\mathbf{r}) = \arctan \frac{LPF(s_1(\mathbf{r}) \times T(\mathbf{r}))}{LPF(s_2(\mathbf{r}) \times T(\mathbf{r}))}$$
(4)

In **Figs. 1-7**, find step-by-step illustration of the presented methodd.

Step 1. Raw data of pajamas with phase slips (**Fig. 1**).

Step 2. Performing a *fast Fourier transform* to determine the spatial frequency **q** (**Fig. 2**).

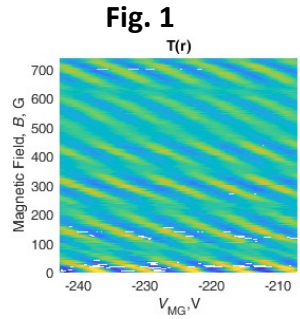

Fig. 1

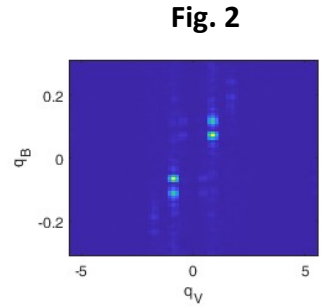

Fig. 2

Since the frequency $q_B$ is not a single sharp peak, the frequency **q** used in the analysis is defined by choosing a frequency between the two visible peaks that best reflects the stepwise behavior of $\theta(\mathbf{r})$.

Step 3. Multiplication of pajama in **Fig. 1** by $s_1$ and $s_2$ (**Fig. 3**, see **Eq. 3**).

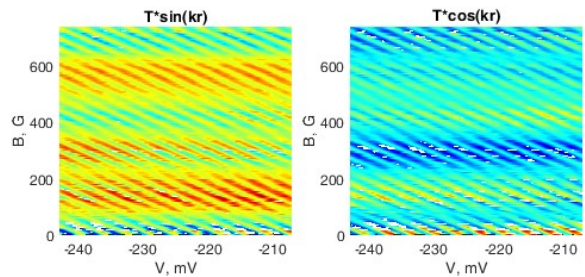

Fig. 3

Step 4. Perform Low-Pass filtering (**Fig. 4, Fig. 5**):

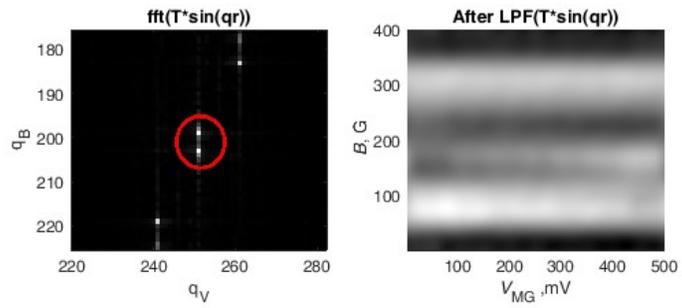

**Fig. 4**

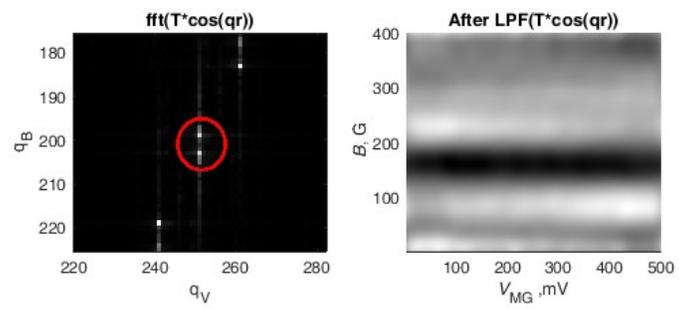

**Fig. 5**

Step 5: Color plot of the $\theta(B, V_{mg})$ (**Fig. 6**)

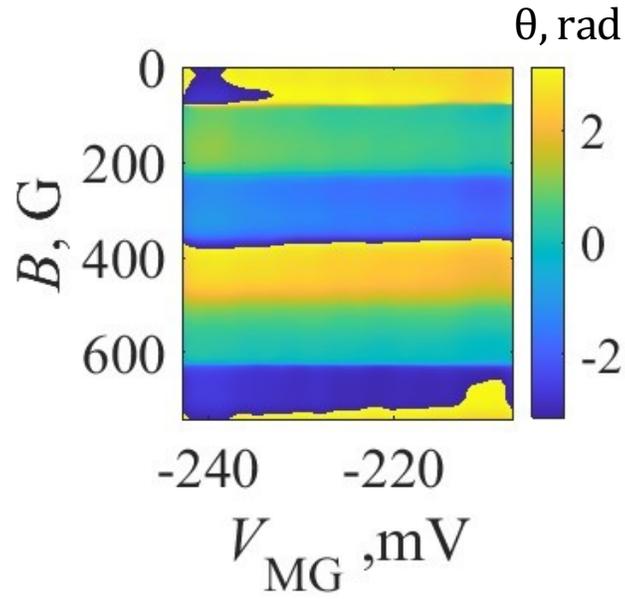

**Fig. 6**

Step 6: Averaging in a range of voltages $V_1 < V < V_2$ (**Fig. 7**, **Table 2**).

$$\nu = 2/5$$

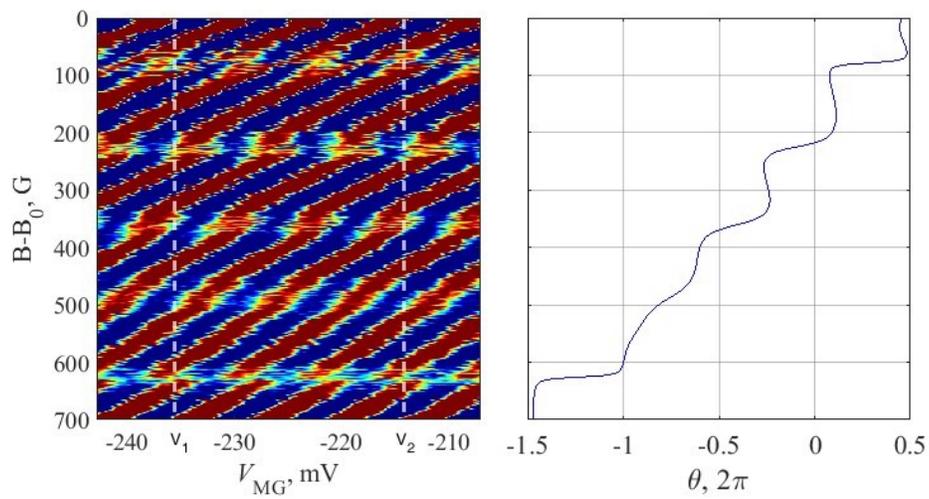

**Fig. 7**

| $V_{MG}, mV$ \ $\nu$ | 1/3 | 2/5 | 3/7 |
|---|---|---|---|
| $V_1, mV$ | $-106.2$ | $-235.6$ | $-465.0$ |
| $V_2, mV$ | $-105.0$ | $-214.2$ | $-455.3$ |

**Table 2**. Voltage range for averaging the phase to get the result in **Fig. 3** from the main text.

The spatial lock-in method is advantageous since it allows isolating of the phase-slips associated with the specific frequency of Aharonov Bohm oscillations. It effectively captures the absolute value of each phase-slip at the chosen frequency. It is less reliable for capturing the direction of the phase slip since the visibility during the phase jumps is significantly suppressed. In addition, the Aharonov-Bohm frequency along the wide range of the magnetic field can be subject to slight change, which may lead to an error in defining the phase flip size.

An alternative method is to use a fast Fourier transform for each horizontal slice (fixed magnetic field) and look at the evolution of the complex phase along the magnetic field. This method might detect the frequency change along $B$ axis more accurately. However, a linear correction is required to determine the phase slip, which is unnecessary in the above mentioned method.

## S4. Interfacing Abelian fractional states

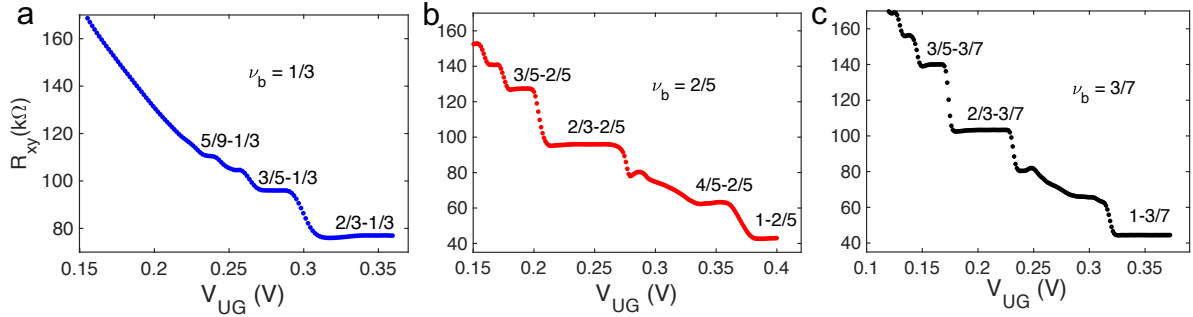

**Figure S2**: **Interfacing Abelian fractional states**. Two terminal resistance measurements at the interface between $\nu_u - \nu_b$. The bulk is fixed at $\nu_b = 1/3, 2/5, 3/7$, and the upper gate (UG) is swept with positive bias. Clear quantized plateaus corresponding to **(a)** $\nu_u$= 5/9, 4/7, 3/5 and 2/3; **(b)** $\nu_u$= 3/5, 2/3, 4/5 and 1; **(c)** $\nu_u$= 3/5, 2/3 – being accurate to ~1%. For $\nu_b = 2/5$, $\nu_u$=1 corresponds to a 150% increase in the 2DEG density underneath the gate. After the charge has equilibrated, the '$\nu_u - \nu_b$' interface charge mode is accompanied by upstream neutral modes.

## S5: QPC response of the OMZI in the integer regime:

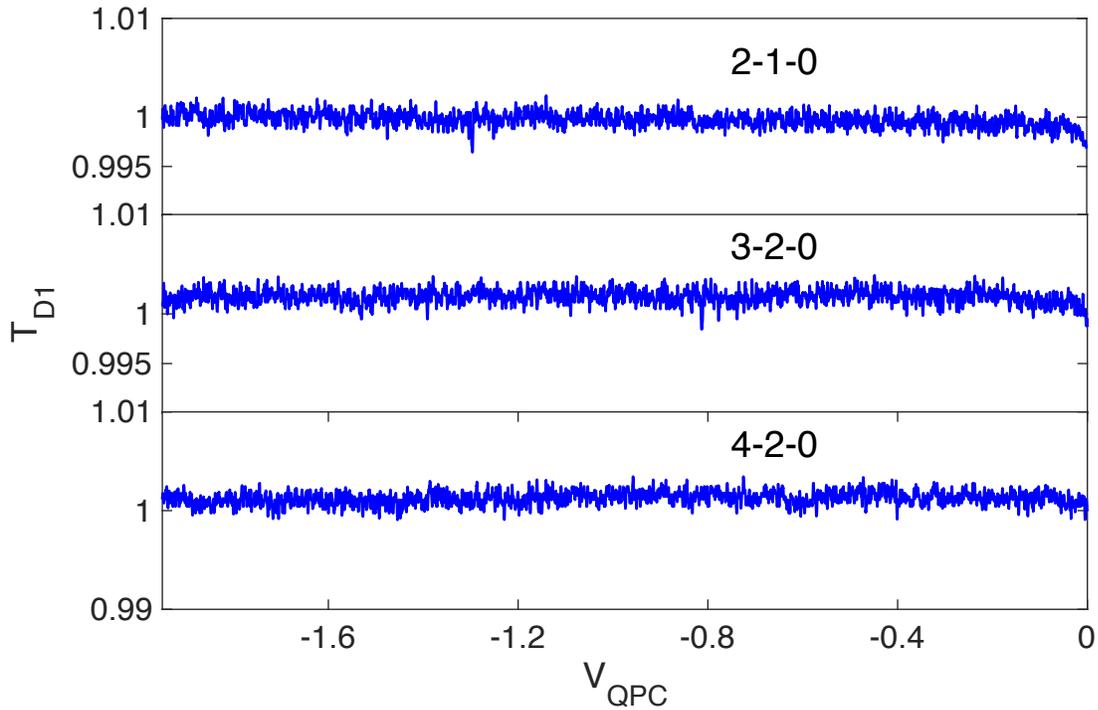

**Figure S3**: **QPC response at integer fillings**. QPC transmission of integer edge modes as a function of QPC's gate voltage for filling factor configurations (a) 2-1-0, (b) 3-2-0, and (c) 4-2-0. As the QPC is pinched, transmission remains unchanged because the opposite spin of the co-propagating channels on opposite sides prevents partitioning at the QPC. The absence of partitioning in the QPC prevents interference of integer edge mode in the OMZI. It is possible that QPC partitioning and interference of integer edge mode in OMZI is feasible at higher filling factors or in lower density samples where the Zeeman splitting is significantly smaller.

## S6. QPC response of the OMZI in the fractional regime

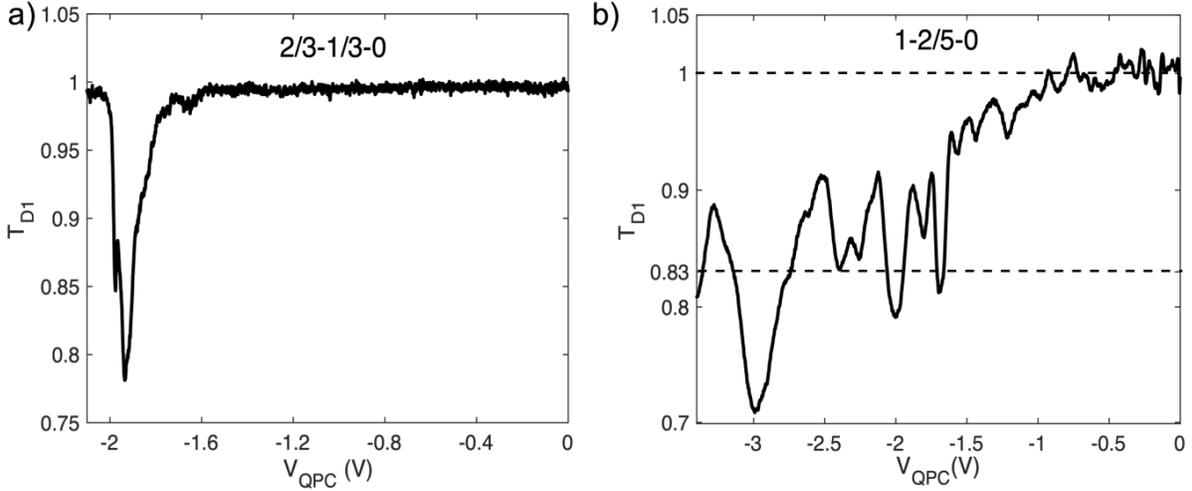

**Figure S4: QPC response in the fractional regime.** One arm of the QPC is negatively biased, while the other arm is positively biased. Due to the chirality, the incident current must pass through the QPC – very different from a conventional QPC. We show the measured QPC transmission as a function of the QPC's gate voltage for filling factor configurations: **(a)** 2/3-1/3-0 and **(b)** 1-2/5-0. In **(b)**, the transmission of $t\sim83\%$ corresponds to the full inner 1/15 mode propagating from S1 to D2 (see Fig.1 of the main text).

## S7: Interference of counter-propagating edge modes

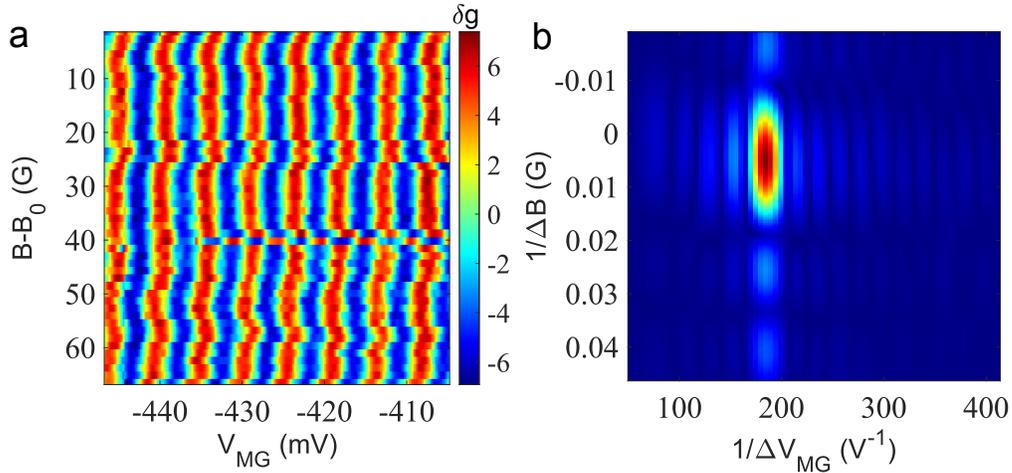

**Figure S5**: **Interference in Fabry-Perot interferometer at $\nu=1/3$. (a)** Transmission oscillations of the edge mode at $\nu=1/3$ in the 0-1/3-0 configuration. While the transmission is independent of the magnetic field $B$, it oscillates as a function of $V_{MG}$. This is the typical behavior of Coulomb-Dominated (CD) interference. **(b)** 2D FFT gives the periodicity $\Delta V_{MG}=5.5$ mV.

## S8: Large-scale pajamas at different positions of 1/3 plateau:

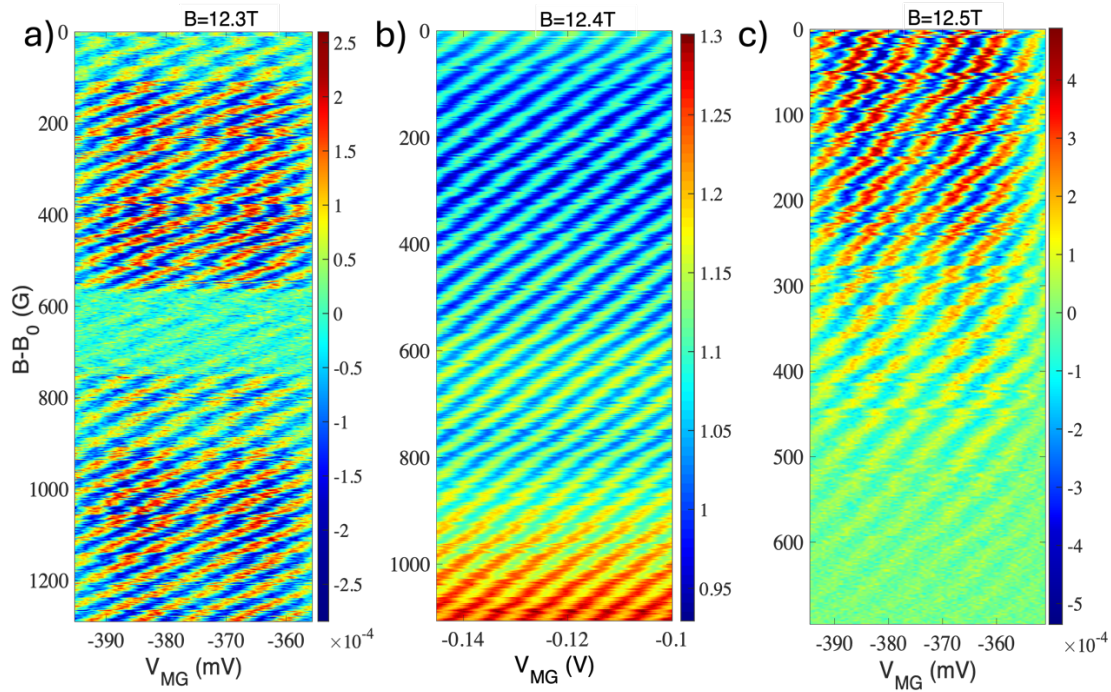

**Figure S6: Aharonov Bohm pajamas of OMZI at different positions of $\nu=1/3$ plateau.** Transmission oscillations as a function of modulation gate voltage $V_{MG}$ and magnetic field B in the filling factor configuration 2/3-1/3-0. The top-gate is fixed at $V_{TG}=0V$. Three AB pajamas (**a,b,c**) were recorded at different positions on the 1/3 plateau starting from the middle of the plateau towards the end at a higher magnetic field, spanning a range of approximately 2,700G. The pajamas are free of discrete phase slips, suggesting *pristine* AB interference in the OMZI.

## S9: Flux periodicity dependence on interface mode on the population side

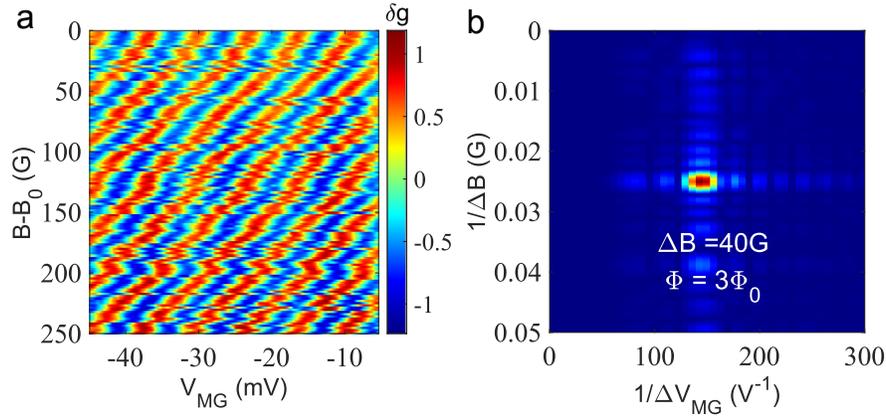

**Figure S7**: **Aharonov Bohm pajama in OMZI at $\nu=1/3$** (see figure above). **(a)** Conductance oscillations with an interfering 1/3 mode in the $B$-$V_{MG}$ plane with the filling-factor configuration 5/9-1/3-0. The top gate is fixed at $V_{TG}=0$ V. As anticipated, the pajama is free from phase flips. **(b)** 2D Fourier transforms of the pajama plot showing a single peak. The periodicity in $B$ is 40G and in $V_{MG}$ is 7mV. The estimated AB area is ~3 µm², leading to a flux periodicity $3\Phi_0$.

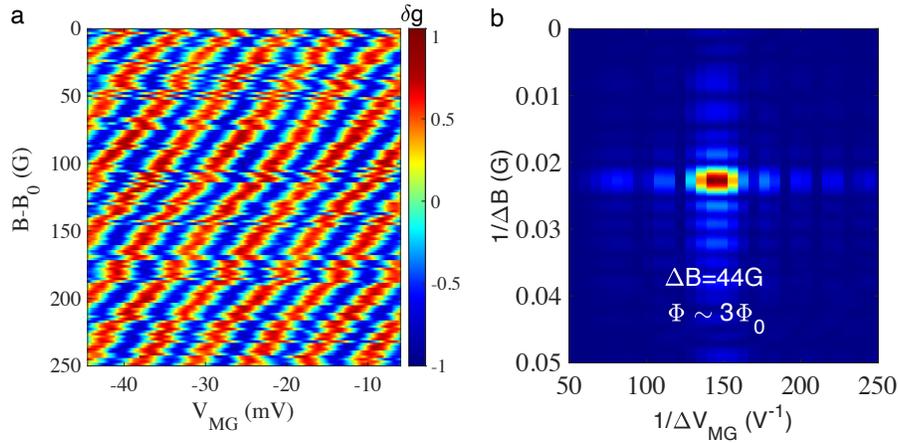

**Figure S8**: **Aharonov Bohm pajama in O MZI at $\nu=1/3$**. **(a)** Transmission oscillations with the interfering 1/3 edge mode in the $B$-$V_{MG}$ plane with a 3/5-1/3-0 configuration. The top gate (of the anti-dot) is fixed at $V_{TG}=0$ V. The Pajama is free of phase jumps. **(b)** 2D Fourier transforms of the pajama plot showing a single peak. The periodicity in B is 44G and in $V_{MG}$ it is 6.8mV. The calculated AB area is ~2.82 µm², leading to a flux periodicity ~$3\Phi_0$. Similar results were observed for '2/3-1/3-0' (see main text), and '5/9-1/3-0' (Figure S7) and for an outer partitioned 1/3 mode in the '1-2/5-0' configuration (Figure S11), suggesting flux periodicity is solely determined by the QPC filling and independent of $\nu_u$. [*the broken pajama lines are du to instabilities*]

## S10. Interference visibility in OMZI:

The visibility of interference oscillations in the OMZI is significantly smaller than in conventional MZI. The low visibility in the OMZI arises due to several factors: i) The interfering edge mode is always accompanied by a neutral mode(s). The neutral mode(s) likely dephase the coherent interference. ii) On the accumulation side, the interface edge mode resides under the metallic gate. Due to the capacitive coupling, the gate can act as a high-pass filter. c) The OMZI is based on the interface edge modes, which are confined to the interface between two bulk regions $'\nu_u - \nu_b'$ and $'\nu_b - \nu_d'$. Lack of sharpening of the edge mode potential might lead to averaging the accumulated AB phase.

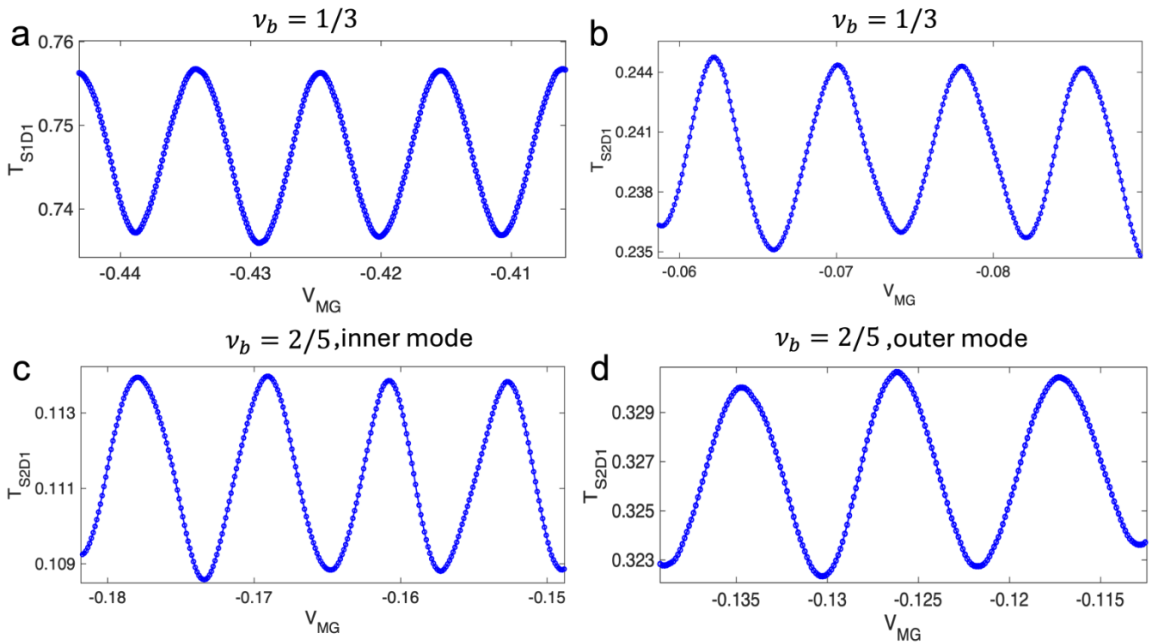

**Figure S9: Visibility calculation in OMZI.** Transmission oscillations as a function of the modulation gate (MG) voltage. The OMZI can be sourced either from the depletion side (from S1) or the accumulation side (from S2). **(a)** Transmission oscillation ($T_{S1D1}$) in the filling factor configuration 2/3-1/3-0, where both source ($S_1$) and detector ($D_1$) reside on the depletion side. The visibility of the oscillation is ~2.8%. **(b)** Oscillations of the transmission ($T_{S2D1}$) in $\nu_b = 1/3$, where the source ($S_2$) is on the accumulation side and the detector ($D_1$) is on the depletion side. The visibility of the oscillation here is ~3.6%. **(c)** Transmission oscillations ($T_{S2D1}$) for the inner 1/15 mode in $\nu_b = 2/5$. The visibility of the oscillations is 4.4%. **(d)** Transmission oscillations ($T_{S2D1}$) for the outer 1/3 mode in $\nu_b = 2/5$. The visibility of the oscillation is 2.7%. The visibility in $\nu_b = 3/7$ varies around ~2%.

## S11: Interference of outer edge modes in an optical-like MZI

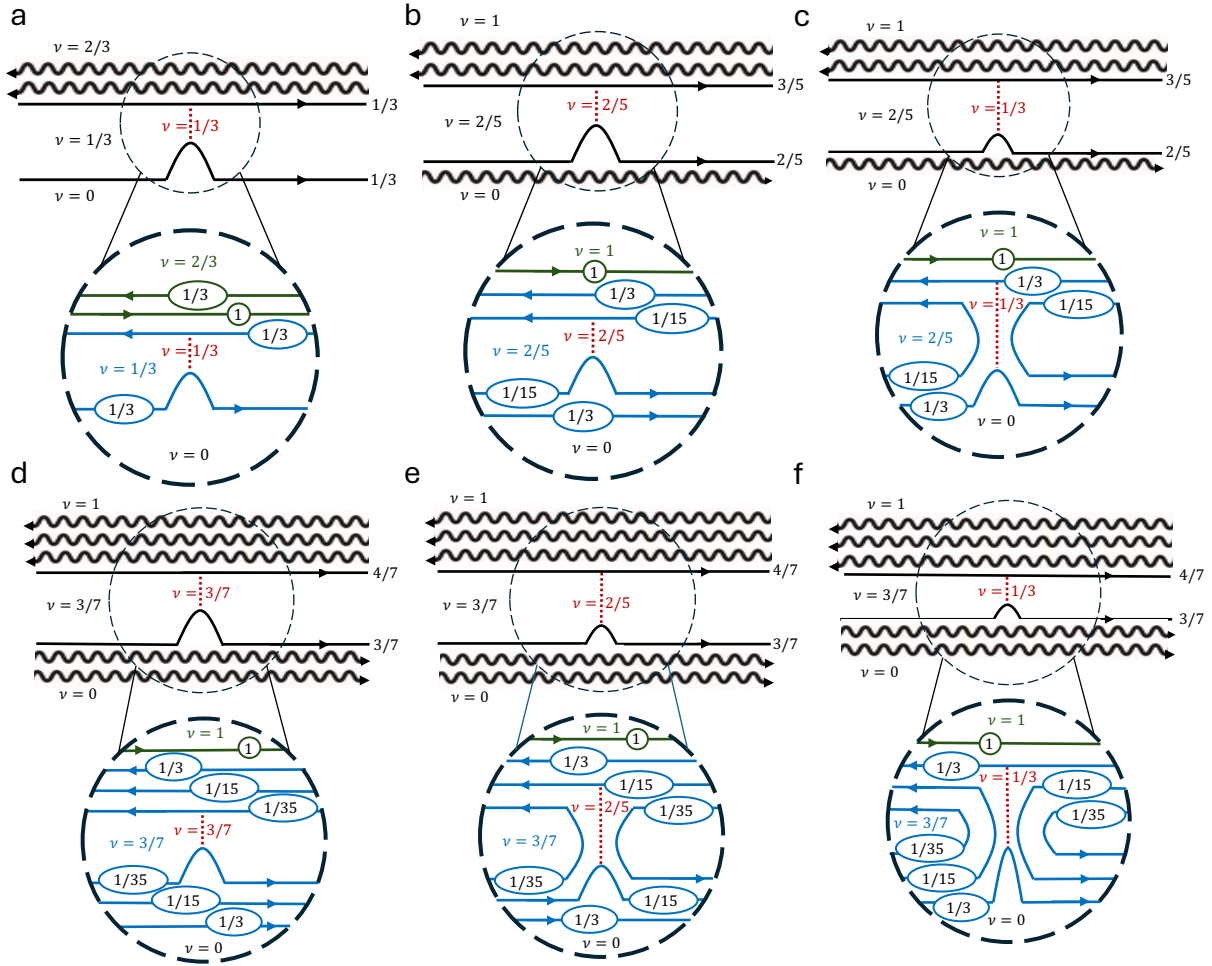

**Figure S10**: **Partitioning of the co-propagating edge modes in the QPC**. The filling factor inside the QPC determines the charge of the tunneling QPs, and thus, the AB flux periodicity and the statistical phase-slips. The solid lines indicate charged modes, and the wavy lines are neutral modes. The equilibrated modes are shown on the top, while the modes before equilibration are shown in the 'zoom-in' below for each combination. The green color indicates the modes corresponding to $\nu_u$, while the blue color indicates the modes corresponding to $\nu_b$. The red number denotes the filling factor inside the QPC. The six shown configurations are: **(a)** Partitioning the 1/3 mode: $\nu_{bulk}$ = 1/3, $\nu_{qpc}$ = 1/3. **(b)** Partitioning the inner 1/15 mode: $\nu_{bulk}$ = 2/5, $\nu_{qpc}$ = 2/5. **(c)** Partitioning the outer 1/3 mode: $\nu_{bulk}$ = 2/5, $\nu_{qpc}$ = 1/3. **(d)** Partitioning the inner 1/35 mode: $\nu_{bulk}$ = 3/7, $\nu_{qpc}$ = 3/7. **(e)** Partitioning the central 1/15 mode: $\nu_{bulk}$ = 3/7, $\nu_{qpc}$ = 2/5. **(f)** Partitioning the outer 1/3 mode: $\nu_{bulk}$ = 3/7, $\nu_{qpc}$ = 1/3.

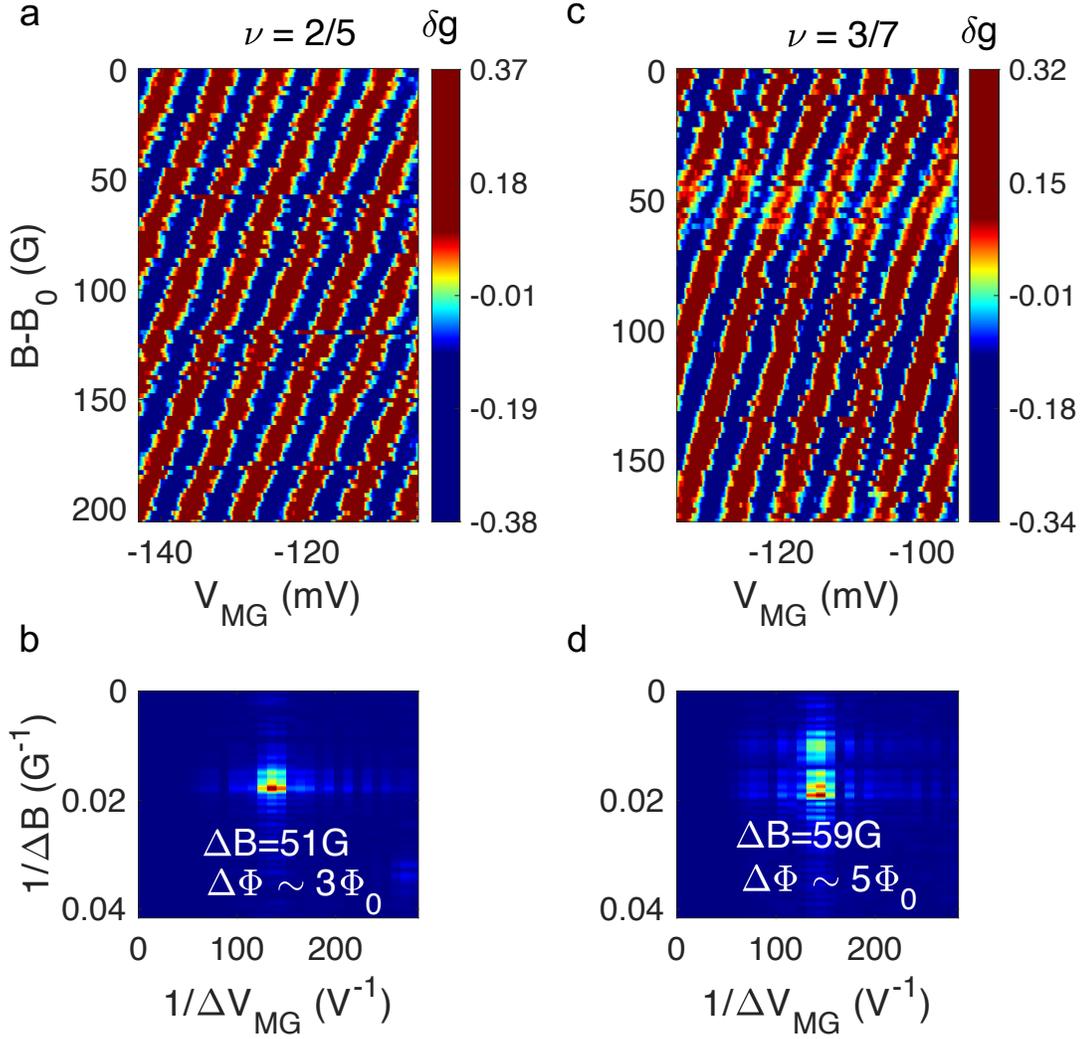

**Figure S11**: **Aharonov Bohm pajamas of the co-propagating outer edge mode with bulk filling $\nu_u$=2/5 and $\nu_u$=3/7.** For these measurements, the QPCs are biased to 90% transmission of the outer 1/3 mode (or the 1/15 mode) of the bulk with filling factor 2/5 (or 3/7). Top gate (TG) is kept at $V_{TG}$=0V. Within the tolerable voltage limit of the QPCs, we could not partition the outer 1/3 mode of the ν=3/7 bulk. The color scale shows conductance oscillations with clear periodicities corresponding to different QPC filling: $3\Phi_0$ (a,b) and $5\Phi_0$ (c,d). The estimated AB area for the outer mode of ν=2/5 is ∼2.45 μm², and for the middle edge of ν=3/7 is ∼3.5 μm². As explained in the main text, the small area correction depends on the pinching voltage of the QPC.

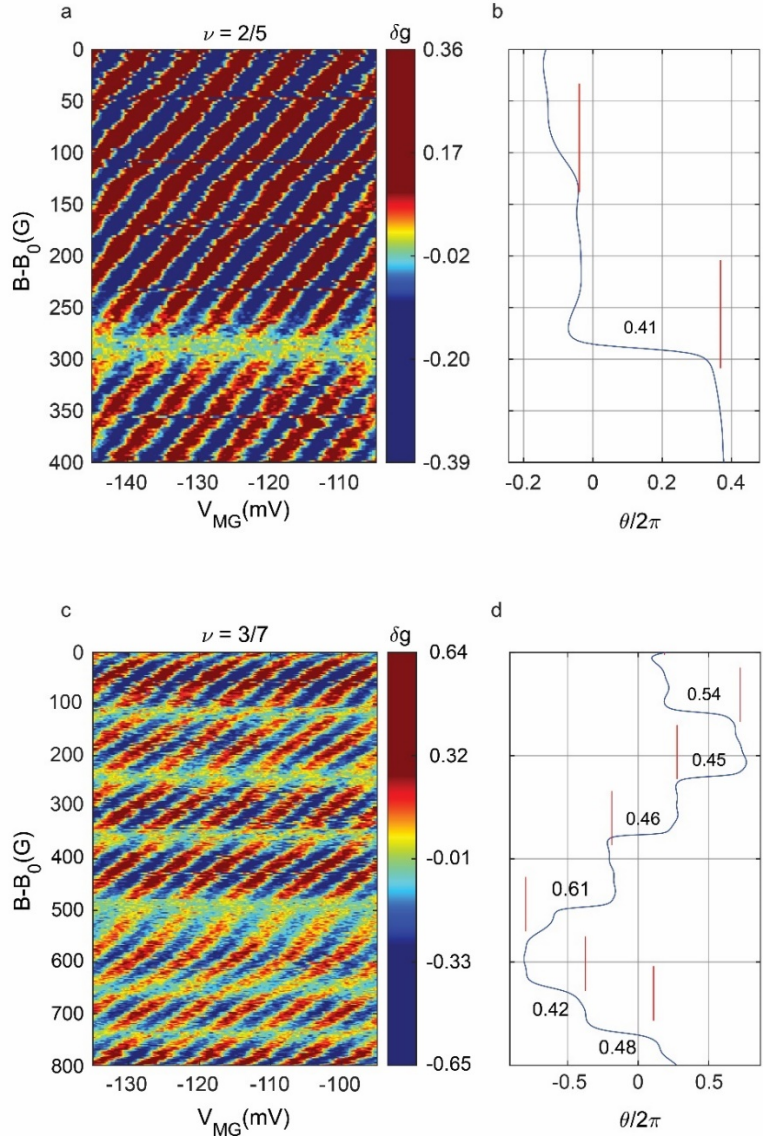

**Figure S12**: **Aharonov Bohm pajamas with phase slips of the co-propagating outer edge mode at bulk filling $\nu_B$=2/5 and 3/7.** The conductance oscillations as a function of B and modulation gate voltage, $V_{MG}$, show AB interference with clear phase-slips. For these measurements, the QPCs are tuned to 90% transmission of the outer filling of 1/3 mode **(a)** [or middle 1/15 mode **(c)** of 2/5 (or 3/7)]. The top-gate voltage $V_{TG}$ is tuned to -50 - 90mV, far from complete depletion under the top-gate. The discrete phase jumps are similar to those observed for the corresponding innermost modes. Figures (**b**) and (**d**) present the calculated phase slips with the Lock-In technique (see **Methods** and **S3**). The red lines indicate the regions of a constant phase.